\documentclass[aps,prd,floatfix,nofootinbib,twocolumn,10pt]{revtex4}

\usepackage{ulem}
\usepackage[utf8]{inputenc}
\usepackage{amsmath}
\usepackage{amssymb}
\usepackage{graphicx,bm}
\usepackage[colorlinks,citecolor=blue]{hyperref}
\usepackage[caption=false]{subfig}
\usepackage{slashed}
\begin{document}
\title{Constraints on dark matter annihilation from the FAST observation of the Coma Berenices dwarf galaxy}
\author{Wen-Qing Guo$^{1,2}$, Yichao Li$^{3}$, Xiaoyuan Huang$^{1,2}$\footnote{Corresponding author: xyhuang@pmo.ac.cn}, Yin-Zhe Ma$^{4,5,6}$\footnote{Corresponding author: ma@ukzn.ac.za}, Geoff Beck$^7$, Yogesh Chandola$^{6}$, Feng Huang$^8$}
\affiliation{$^{1}$Key Laboratory of Dark Matter and Space Astronomy, Purple Mountain Observatory, Chinese Academy of Sciences, Nanjing 210023, China \\
$^{2}$School of Astronomy and Space Science, University of Science and Technology of China, Hefei, Anhui 230026, China \\
$^3$Department of Physics, College of Sciences, Northeastern University, Shenyang 110819, China \\ 
$^{4}$School of Chemistry and Physics, University of KwaZulu-Natal, Westville Campus, Durban, 4000, South Africa \\
$^{5}$NAOC-UKZN Computational Astrophysics Centre (NUCAC), University of KwaZulu-Natal, Durban, 4000, South Africa \\
$^{6}$Key Laboratory of Radio Astronomy, Purple Mountain Observatory, Chinese Academy of Sciences, 
Nanjing 210023, China\\
$^7$School of Physics and Centre for Astrophysics, University of the Witwatersrand,
Johannesburg, Wits 2050, South Africa \\
$^8$Department of Astronomy, Xiamen University, Xiamen, Fujian 361005, China
}

\begin{abstract}
The Galactic center $\gamma$-ray excess, detected  by the Fermi-LAT, is a very attractive tentative signal from dark matter annihilation. Searching for associated synchrotron emissions can test the dark matter interpretation for this excess. We geared the Five-hundred-meter Aperture Spherical radio Telescope (FAST) towards Coma Berenices (a dwarf Spheroidal galaxy) for 2-hours of observation, and found no significant continuum radio emission, which could { put constraints on the dark matter annihilation cross section}, from our target. We set very stringent annihilation cross-section constraints, with roughly an order of magnitude improvement over a large range of masses compared with previous radio searches. The dark matter scenario for the Galactic center $\gamma$-ray excess is in tension with the FAST observation for reasonable choices of astrophysical factors. {But considering the large  uncertainty on astrophysical parameters, such as the magnetic field, the diffusion coefficient, and the diffusion radius, and on DM halo parameters, the dark matter interpretation for the excess could still survive.} Further radio observations by the FAST and other radio telescopes may reveal more about the dark matter properties. 
\end{abstract}

\date{\today}

\maketitle

\section{Introduction}
Observations from galaxy rotation curves, weak gravitational lensing, and anisotropies in the Cosmic Microwave Background (CMB) coherently suggest the existence of dark matter (DM) in the Universe, which constitutes about 26\% of the cosmic energy budget~\cite{Planck:2018vyg}. Though the fundamental nature of DM is still a mystery, several well-motivated candidates from particle physics have been proposed~\cite{Bertone:2004pz}. One plausible and widely accepted candidate is the Weakly Interacting Massive Particle (WIMP), whose efficient self-annihilation in the early Universe can naturally explain the observed relic density today~\cite{Steigman:2012nb}. 

WIMPs can produce Standard Model (SM) particles through self-annihilation. Thus, searching for signals of these particles, and their secondary emissions, could reveal the nature of dark matter. This is one of the most promising DM search methods~\cite{Bertone:2004pz}. In the past decade, there have been several astrophysical anomalies reported and investigated, which could be potential DM annihilation signature. These include the positron and electron excesses ~\cite{PAMELA:2008gwm,DAMPE:2017fbg, AMS:2021nhj}, the antiproton excess \cite{Cui:2016ppb,Cuoco:2016eej}, and the Galactic Center $\gamma$-ray excess (GCE; ~\cite{Hooper:2010mq,Zhou:2014lva,Daylan:2014rsa,Calore:2014xka,Huang:2015rlu}). The GCE, detected by Fermi Gamma-Ray Space Telescope, is peaked at energies $\sim 1-3$ GeV and extends out to at least 10$^\circ$ from the GC ~\cite{Calore:2014xka, Huang:2015rlu}. Several works have found that its spectrum is consistent with the signal of WIMP annihilating to $b\bar{b}$ or $\tau^{+}\tau^{-}$ with the mass of $\sim \mathcal{O}(10)\,{\rm GeV}$, and its morphology is broadly consistent with the predicted signal of dark matter annihilation ~\cite{Calore:2014xka,Zhou:2014lva, Daylan:2014rsa, Huang:2015rlu,DiMauro:2021raz}. In addition, it has been shown that the channel of WIMP annihilating into $\mu^+\mu^-$ with the associated IC emission, can fit the flux spectrum, in consistence with current constraints from the combined dwarf spheroidal galaxies analysis and the AMS-02 $\bar{p}$ and e$^+$ data~\cite{DiMauro:2021qcf}. The possible connections
between the DM interpretation of the GCE and other anomalies have also been investigated~\cite{Fan:2022dck,Zhu:2022tpr}. Apart from DM, several astrophysical models were proposed for this excess, such as  cosmic rays ~\cite{Carlson:2014cwa,Gaggero:2015nsa,Cholis:2015dea,Carlson:2016iis}, and unresolved millisecond pulsars ~\cite{Yuan:2014rca,Bartels:2015aea,Lee:2015fea,Macias:2016nev,Bartels:2017vsx,Gautam:2021wqn}. Though more sophisticated methods are employed in the  analysis, it is still difficult to discriminate the scenarios of DM annihilation and unresolved millisecond pulsars solely by the gamma-ray observation~\cite{Leane:2019xiy, Zhong:2019ycb,Calore:2021jvg,Cholis:2021rpp}.  Therefore, multiwavelength observation is essential to test the nature of the GCE.

Electrons and positrons, produced by DM annihilation, can lose their energies by Inverse Compton Scattering (ICS) and synchrotron radiation, which generate fluxes in X-rays and radio wavelength respectively, with detailed variation determined by the DM mass and the astrophysical environment. If DM is heavy enough to produce electrons, the synchrotron emission can fall into the detectable range of current or future radio telescopes \cite{Colafrancesco:2006he,Spekkens:2013ik,Natarajan:2013dsa,Natarajan:2015hma, Colafrancesco:2014coa,Beck:2015rna,Regis:2017oet,Kar:2019hnj,Vollmann:2019boa, Bhattacharjee:2019jce, Bhattacharjee:2020phk, Kar:2020coz, Chen:2021rea, Regis:2021glv,Basu:2021zfg}. Considering the DM annihilation scenario for the GCE, all the $b\bar{b}$, $\tau^{+}\tau^{-}$, and $\mu^{+}\mu^{-}$ final states could also generate electrons and positrons, other than $\gamma$-ray emissions. Thus it is possible to search for associated synchrotron emissions from these charged particles and test the DM interpretation. Furthermore, although it is very difficult to probe DM annihilation through muon decays into $\gamma$-rays, decaying muon can produce an abundance of relativistic electrons, whose emissions may instead be detectable via radio observations.

In this work, we perform the DM indirect search with the Five-hundred-meter Aperture Spherical radio Telescope. The telescope is the largest filled-aperture, single-dish, radio antenna in the world, with a very high sensitivity~\cite{Jiang:2019rnj,2020RAA....20...64J}, and its 19-beam receiver covers the frequency range of 1-1.50 GHz \cite{Nan:2011um,2020RAA....20...64J, 2020Innov...100053Q}. FAST is located at the geographic latitude of $25^{\circ}29'10''$ and its observable maximum zenith angle is $40^{\circ}$, covering a large portion of the sky but without the GC region. Thus instead of the GC, we choose the Coma Berenices dwarf galaxy as our target to constrain the WIMPs in this paper, considering its location on the sky and its proximity.

The paper is organised as follows. In Section~\ref{Target and Expected Signal}, we illustrate reasons of the Coma Berenices selection, and calculate the expected signal from DM annihilation. In Section~\ref{Results}, we conduct the FAST data analysis and place upper limits  of the WIMP parameter in $\mu^{+}\mu^{-}$, $\tau^{+}\tau^{-}$ and $b\bar b$ final states, and further discuss the results and the GCE. We conclude in Section~\ref{conclusion}.

\section{Target and Expected Signal}\label{Target and Expected Signal}

\subsection{Target selection}
The GC, where the GCE is observed, is not in the field of view of the FAST. However, since DM is a ubiquitous in cosmic structure, emissions within other targets should be consistent with DM signals in the GC. Dwarf Spheroidal galaxies (dSphs) of the Local Group are considered to be one of the best targets to search for annihilation signals from DM particles. These satellite galaxies of the Milky Way (MW) are located within ${\cal O}$ (100 kpc) { of MW center } and are basically free of radio background since they are virtually empty of gas, dust, and recent star formation. Furthermore, their mass-to-light ratios can be as high as $\sim \mathcal{O}(100)$~\cite{Fermi-LAT:2010cni}, among the highest in the Universe~\cite{Strigari:2007at}. From $\gamma$-ray observations, these dSphs have provided tight constraints on the dark matter interpretation of the GCE \cite{Fermi-LAT:2016uux,Keeley:2017fbz,DiMauro:2021qcf}. 

Coma Berenices was discovered by the Sloan Digital Sky Survey~\cite{SDSS:2006fpg}, located at a distance of about 44 kpc from the Sun, centered at $\rm RA = 12h26m59s$, $\rm DEC = +23^{\circ} 54'00''$~\cite{Kalashev:2020hqc} in equatorial coordinates (J2000). Coma Berenices has a high mass-to-light ratio~\cite{Simon:2007dq}, known as the hitherto darkest galaxies, which can be expected to have a strong DM signal~\cite{Bonnivard:2015xpq}. In addition, the Coma Berenices galaxy is almost unaffected by tidal disruption from the Milky Way~\cite{Munoz:2009hj}. {Therefore, the stellar kinematic properties of this dSph are determined mainly by the gravitational potential of its DM halo}. More importantly, the position of the Coma Berenices falls within the proper zenith angle suitable for the FAST observation~\cite{2020RAA....20...64J}. In this paper, we use the FAST observation of Coma Berenices to search for the DM annihilation signals.

\subsection{Expected signal}
Charged particles propagating through the interstellar medium can lose energy owing to a variety of radiative processes, such as inverse Compton radiation, synchrotron radiation, Coulomb losses and bremsstrahlung. The radiation can be substantial only if the particle is sufficiently long-lived and the mass is small enough. Therefore, electrons and positrons produced by DM annihilation can generate strong radiation.  In this work, we employ the  publicly available package {\tt RX\-DMFIT}~\cite{McDaniel:2017ppt} to calculate the expected signal from DM annihilation. 

Several dark matter halo profiles have been investigated in numerical simulations, and observed nearby dSphs prefer flattened cores rather than the cusps predicted by N-body simulations~\cite{Burkert:1995yz}, with some exceptions among the ultra-faint satellites~\cite{Hayashi:2022wnw}. It is thought that such cores may result from stellar feedback during the galaxy formation \cite{Maccio:2011ryn}. In this paper, we take the  Einasto DM profile instead of the cuspy profile to describe the DM halo of Coma Berenices
\begin{equation}
\label{Einasto profile}
\rho_{\rm \chi}(r)=\rho_{0}\exp\left[-\frac{2}{\alpha}\left(\left(\frac{r}{r_{s}}\right)^{\alpha} - 1\right)\right],
\end{equation}
where $\alpha=0.847$, the scale radius $r_{\rm s}=5.14\,{\rm kpc}$ and the normalisation $\rho_{0}=1.962\,{\rm GeV}\,{\rm cm}^{-3}$, which are best-fitting values derived from stellar kinematic data  \cite{Bonnivard:2015xpq}. 

To calculate the synchrotron emission from dark matter annihilation, we need to solve the diffusion equation, with a steady state solution, to obtain the equilibrium $e^{\pm}$ spectrum. Here, we assume the spherically symmetric case and calculate
\begin{eqnarray}
\label{diffusion equation}
\frac{\partial}{\partial t} \frac{\partial n_{\rm e}}{\partial E}
&=& \nabla\cdot\left[D(E,r)\nabla\left(\frac{\partial n_{\rm e}}{\partial E}\right)\right] \nonumber \\
&+& \frac{\partial}{\partial E}\left[b(E,r)\frac{\partial n_{\rm e}}{\partial E}\right]+Q(E,r),
\end{eqnarray}
where $\partial n_{\rm e}/\partial E$ is the differential equilibrium electron density, $D(E,r)$ is the diffusion coefficient, $ b(E,r)$ is the energy loss term, and $Q(E,r)$ is the source term of electron. The source term $Q(E,r)$ is assumed to be entirely determined by DM annihilation, which is given by,
\begin{equation}
\label{annihilation term}
Q_{\rm ann}(E,r)=\frac{\langle\sigma v\rangle\rho^{2}_{\chi}(r)}{2M^{2}_{\chi}}\sum_{f}{\rm BR}_{f} \frac{{\rm d}N^f}{{\rm d}E}(M_{\chi}),
\end{equation}
where $\langle\sigma v \rangle$ is the velocity averaged total annihilation cross-section, $\rho_{\chi}(r)$ describes the DM density distribution in the dwarf galaxy, $M_{\chi}$ is the mass of the DM particle, ${\rm BR}_{f}$ is branching ratio to final state $f$, and ${\rm d}N^f/{\rm d} E$ is the electron injection spectrum through channels $f$ determined by the public package $\rm{DarkSUSY}$ ~\cite{Bringmann:2018lay}. 

For the diffusion coefficient, in the absence of detailed knowledge of the structure of the dSphs, we adopt a spatially independent form with a power-law energy dependence as 
\begin{equation}\label{diffusion coefficient}
    D(E)=D_{0}\left(\frac{E}{1\,\rm GeV}\right)^{\gamma},
\end{equation}
where $D_{0}$ is the diffusion constant, which default value is chosen to be $10^{27}\,{\rm cm}^{2}\,{\rm s}^{-1}$ to obtain the estimate for the radio signal~\cite{Jeltema:2008ax,Vollmann:2019boa,Kar:2020coz}. Of course, in Fig.~\ref{fig:mu_discuss} we vary this parameter to see how the constraint may change. $\gamma$ is chosen to be $1/3$, consistent with the observations in the Milky Way \cite{Korsmeier:2016kha}.

In Eq.~(\ref{diffusion equation}), the full energy loss term has contributions from synchrotron, Inverse Compton Scattering, Coulomb, and bremsstrahlung losses. Each energy loss term depends on the energies of the electrons and positrons, and the magnetic field strength for synchrotron losses and the photon spectrum for ICS losses. The full energy loss equation is 
\begin{eqnarray}
b(E, r) &=& b_{\rm IC}(E)+b_{\rm syn}(E, r)+b_{\rm coul}(E) \nonumber \\
&+& b_{\rm brem}(E) \nonumber \\
&=& b_{\rm IC}^{0} \left(\frac{E}{1\,\rm GeV}\right)^{2}+b_{\rm syn}^{0} \left(\frac{B_{0}}{1\,\rm \mu G}\right)^{2} \left(\frac{E}{1\,\rm GeV}\right)^{2} \nonumber \\
&+& 
b_{\text {coul}}^{0} n_{\rm e}\left[1+\log \left(\frac{E / m_{\rm e}}{n_{\rm e}}\right) / 75\right] \nonumber \\
& + & b_{\rm brem}^{0} n_{\rm e}\left[\log \left(\frac{E / m_{\rm e}}{n_{\rm e}}\right)+0.36\right], 
\end{eqnarray}
where $B_{0}$ is the strength of the magnetic field, $n_{\rm e}\approx 10^{-6}$ is the average number density of thermal electrons in dwarf galaxies~\cite{Colafrancesco:2006he, McDaniel:2017ppt}. For Coma Berenices, we assume a uniform magnetic field model in this work as a conservative estimation, i.e. $B_{0} = 1\,\mu{\rm G}$. We take these energy loss rate parameters $b^{0}_{\rm IC}$, $b^0_{\rm syn}$, $b^{0}_{\rm coul}$ and $b^{0}_{\rm brem}$ as 0.25, 0.0254, 6.13 and 1.51 respectively in units of $10^{-16}\,{\rm GeV}\, {\rm s}^{-1}$~\cite{2011hea..book.....L,McDaniel:2017ppt}.

We let Eq.~(\ref{diffusion equation}) equal to zero to seek for a steady-state solution. For the boundary condition, the electron density is generally assumed to vanish at the boundary of the diffusive zone, then the equilibrium electron density can be solved as,
\begin{equation}
  \frac{\partial n_{\rm e}}{\partial E}=\frac{1}{b(E,r)}
  \int^{M_\chi}_{E}  {\rm d}E'\,
  G\left[r, \lambda(E,E',r)\right]
  Q(E',r),
  \label{eq:dndeGreen}
\end{equation}
where the Green's function is obtained by 
\begin{eqnarray}
G(r, \lambda) &=& \frac{1}{\sqrt{4\pi\lambda^{2}}}\sum^{\infty}_{i=-\infty}(-1)^{i}\int^{r_{h}}_{0}{\rm d}r'\left(\frac{r'}{r_{i}}\right) \nonumber \\
& \times &
\left(\frac{\rho_{\chi}(r')}{\rho_{\chi}(r)}\right)^{2} 
\left[\exp\left(-\frac{\left(r'-r_{i} \right)^{2}}{4\lambda^{2}} \right) \right. \nonumber \\
&-& \left.\exp\left(-\frac{\left(r'+r_{i} \right)^{2}}{4\lambda^{2}} \right)\right].
\end{eqnarray}
The summation over $i$ is to take into account the image charges positioned at $r_{i}=(-1)^{i}r+2ir_{\rm h}$, and the $r_{\rm h}$ is the size of the dSphs' diffusion zone,  which is set to be twice the distance of the outermost star of Coma Berenices, $r_{\rm max}=$ 238 pc~\cite{Bhattacharjee:2019jce,Bhattacharjee:2020phk,Geringer-Sameth:2014yza}. $\lambda$ is the mean ``energy-loss'' distance traveled by a charged particle with the source energy $E'$ and the interaction energy $E$:
\begin{equation}
\lambda^2(E,E',r)=\int^{E'}_E \frac{D(\epsilon)}{b(\epsilon,r)} {\rm d}\epsilon .
\end{equation}

Once we obtain the equilibrium electron density, we can calculate the  integrated flux density spectrum for synchrotron emission as
\begin{equation}
    S_{\nu}=2\int_{\hat{\Omega}}{\rm d}\Omega \int_{\rm LoS}\frac{{\rm d}l}{4\pi}\int_{m_{e}}^{M_{\chi}}{\rm d}E\, \mathcal{P}_{\rm syn}(E) \, \frac{\partial n_{\rm e}}{\partial E},
\end{equation}
where the factor 2 takes into account the $e^{\pm}$ pair. $\mathcal{P}_{\rm syn}$ is the standard power of synchrotron emission, which calculation can be found in Ref.~\cite{McDaniel:2017ppt}. The integration radius in this work is $3'$, which is the FWHM of the FAST central beam (Beam 01)~\cite{2020RAA....20...64J}.

\begin{figure}
    \includegraphics[width=1.\columnwidth]{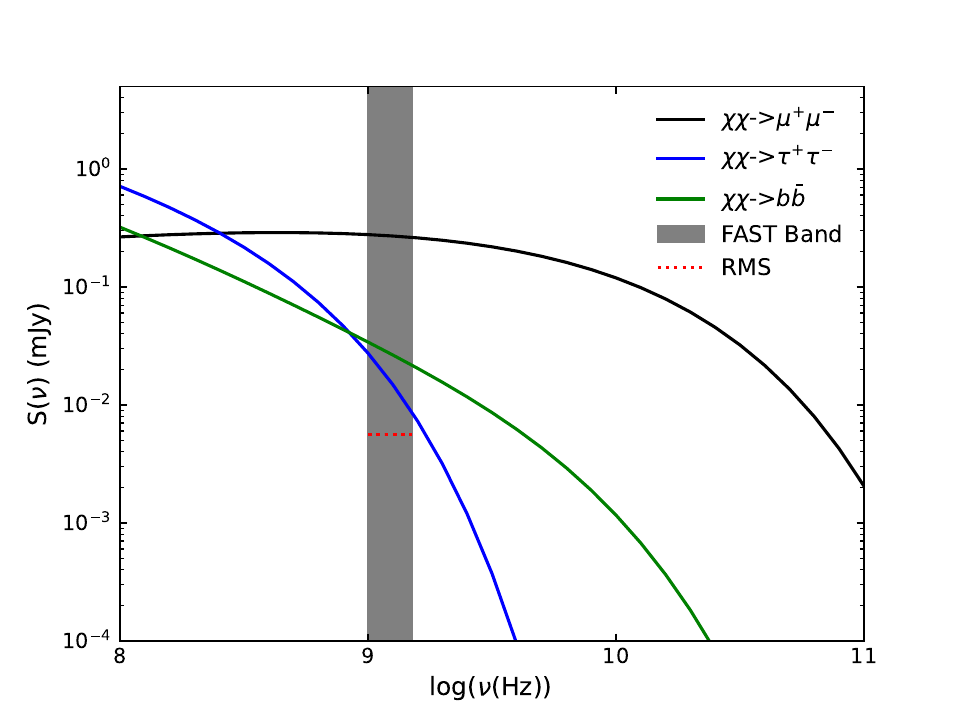}
    \caption{Expected radio emissions from Coma Berenices with DM parameters favored by the GCE. The grey band represents the FAST frequency coverage, ranging from $1,000$-$1,500\,{\rm MHz}$. And the red dotted line shows the RMS value of 1 min observations toward Coma Berenices with the bandwidth 500 MHz.}
    \label{fig:bestfit_flux}
\end{figure}

\setlength{\tabcolsep}{10pt}
\begin{table}[tbp]
\centering
    \begin{tabular}{ccc}
      \hline
      \hline
      Channel                  & $M_{\chi} \rm (GeV)$             & $\langle \sigma v \rangle \rm (\times 10^{-26}cm^{3}\, s^{-1})$ \\
      \hline
      $\mu^{+}\mu^{-}$              & 58         & 3.9 \\
      $\tau^{+}\tau^{-}$            & 7.2        & 0.43 \\
      $b\bar{b}$                    & 42         & 1.41 \\
      \hline
     \end{tabular}
    \caption{The GCE best-fitting value for the DM parameters, $M_{\chi}$ and $\langle \sigma v \rangle$ ~\cite{DiMauro:2021qcf}. }
    \label{GCE best-fitting values}
\end{table}

With these assumptions, we show the expected radio emissions from Coma Berenices with DM parameters favored by the GCE (Table~\ref{GCE best-fitting values}), in Fig.~\ref{fig:bestfit_flux}.  Taking the bandwidth, $\Delta B=500\,{\rm MHz}$, and the integration time, $\tau=1\,{\rm min}$, we can estimate the expected RMS of the FAST observation, via
\begin{equation}
    \sigma_{\rm T}=\frac{T_{\rm sys}}{\sqrt{2\Delta B \tau}},
    \label{equ:rms}
\end{equation}
where $T_{\rm sys}=24\,{\rm K}$~\cite{2020RAA....20...64J}. We also show the band coverage and RMS of the FAST in Fig.~\ref{fig:bestfit_flux}. One can see that FAST RMS noise is lower than the associated radio signal of the GCE, especially for the $\mu^+\mu^-$ channel.

\section{Data Analysis and Results}\label{Results}

\subsection{Observation}
The FAST observation was conducted between 2020-12-14 07:00:00 and 2020-12-14 08:50:00.
The 19-beam receiver equipped on the FAST can record data of two polarisations through 65,536 spectral channels, covering the frequency range of $1,000$-$1,500$\, MHz. We employ the SPEC(W+N) backend in the observation, with 1 second sampling time. To reduce the effects of contamination from the Earth's atmosphere, baseline variation and other foregrounds, we chose the  ``ON/OFF'' observation mode of the FAST. In each round of observation, the central beam of the FAST is pointed to Coma Berenices for 350 seconds, and the central beam of the FAST is moved to the ``OFF source'' position, which is half a degree away from the ``ON source'' without known radio sources, for another 350 seconds. The total time of observation is two hours, comprising of 9 rounds of ON/OFF observation. Here, we utilize the first 6 rounds of observations with stable data taking. We also employ the low noise injection mode for signal calibration, with the characteristic noise temperature about 1.1 K~\cite{2020RAA....20...64J}. In the whole observation, the noise diode is continuously switched on and off with a period of 2 seconds.   

\subsection{FAST data analysis}

\begin{figure*}
    \includegraphics[width=2.\columnwidth]{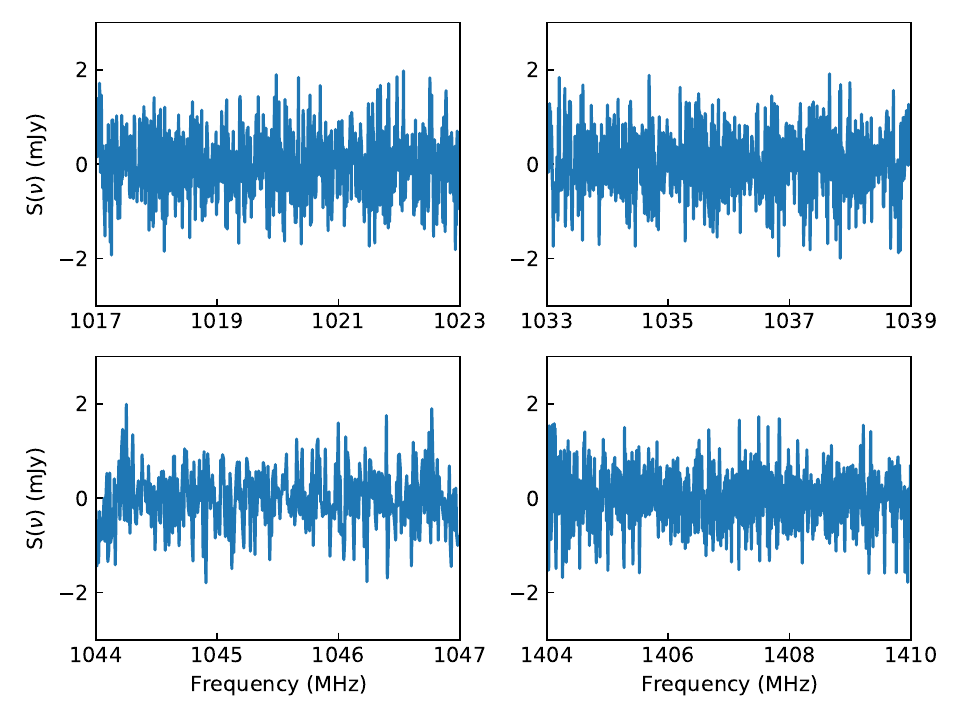}
    \caption{The flux density, converted from source temperature. Different panels show the flux density in four bands of frequencies. For the selected frequency bands, we have applied the data-cleaning process and the periodic structure removal.}
    \label{fig:Density_flux}
\end{figure*}

Let $P^{\rm cal/off}$ be the original instrument reading without noise injection, and let $P^{\rm cal/on}$ be the reading with noise injected. Then the calibrated antenna temperature $T_{\rm A}$ can be derived as~\cite{ONeil:2002amr,2020RAA....20...64J,2021MNRAS.503.5385Z}
\begin{equation}
\label{Temperature calibration}
    T_{\rm A}=\frac{P^{\rm cal/off}}{P^{\rm cal/on}-P^{\rm cal/off}}T_{\rm noise},
\end{equation}
where $T_{\rm noise}$ is the pre-determined noise temperature measured with hot loads. $T_{\rm A}$ from both polarizations were checked for without large variation, and each pair of polarization temperatures was added together to obtain the average. With these procedures, the temperature for each time bin and each frequency bin of the central beam could be obtained. Then, these temperatures in the time domain are checked, and time bins, when the telescope is not stable due to switching between ON and OFF positions, were masked, leaving 1488 time bins (744 time bins for the ON observations and 744 time bins for the OFF observations) for each frequency. The contribution from the background is removed by subtracting $T_{\rm OFF}$ from  $T_{\rm ON}$, yielding source temperature $T$, but this procedure would also remove the DM signal in the OFF region from that in the ON region.  We then convert temperature $T$ to flux density with pre-measured antenna gain, which depends on beam and frequency ~\cite{2020RAA....20...64J}. During the data recording, large noise fluctuations, such as  radio frequency interference (RFI), may be very strong to cover possible DM signals. In order to reduce the effect of these large noise fluctuations, we adopt a data cleaning process to select good data ~\cite{An:2022hhb}. For each frequency bin $i$, we divided 744 data into 28 groups. The first 27 groups contain 27 time bins and the last group contains 15 time bins. We select the group with the smallest variance, $\sigma^{2}_{\rm ref}$, as the reference group. The mean value of the reference group is denoted as $\mu_{\rm ref}$. We then retain the good data with deviation from the reference mean smaller than $5\sigma_{\rm ref}$, i.e. $|D-\mu_{\rm ref}|<5\sigma_{\rm ref}$, where $D$ is the data along the time bins. Finally, we average out time in each frequency bin to get the flux density $S_i$ and we also get the standard deviation $\sigma_i$.

During the FAST commissioning phase, the data are also suffered from a frequency ripple which associated 
with a standing wave between the FAST feed and reflector. The wavelength of the ripple is about $1.1\,{\rm MHz}$. 
Such a frequency ripple significantly increases the variance across frequency channels and can be efficiently 
suppressed with a band-reject filter \cite{2021RAA....21...59L}. 
We build such a filter using a window function in delay-space, which is the Fourier transfer of the frequency space. 
The window function suppresses the Fourier mode amplitude with wavenumber centering at the ripple wavelength. 
The width of the window function is chosen to 
cover from $1/\Delta B_{\rm sub}$ to $1/(0.5\,{\rm MHz})$, where $\Delta B_{\rm sub}$ is the bandwidth of the sub-band. 
With such a filter, the data mean across the frequency, as well as the fluctuations with a wavelength shorter
than $0.5$ MHz, are reserved. The spectroscopic structures removed with such a filter are assumed to be
different from the dark matter signals.
The filter is applied to four sub-bands of the data which are relatively RFI-free and the ripple-removed data, 
denoted as $S_{\rm data}$, are shown in Fig.~\ref{fig:Density_flux}. 

\begin{figure*}[htb]
    \centerline{
    \includegraphics[width=1\columnwidth]{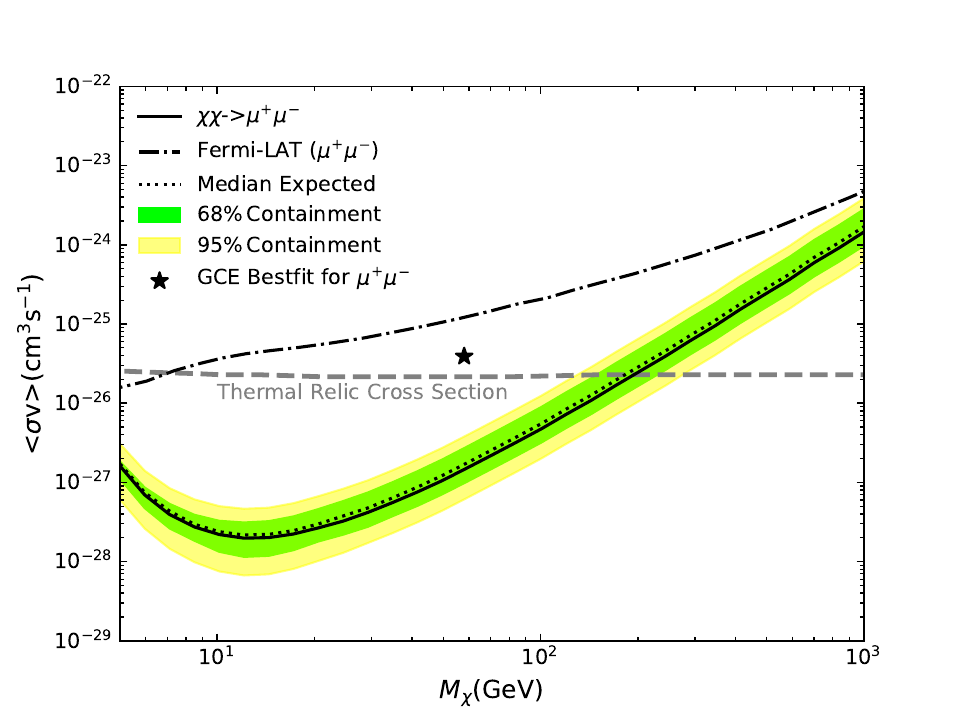}
    \includegraphics[width=1\columnwidth]{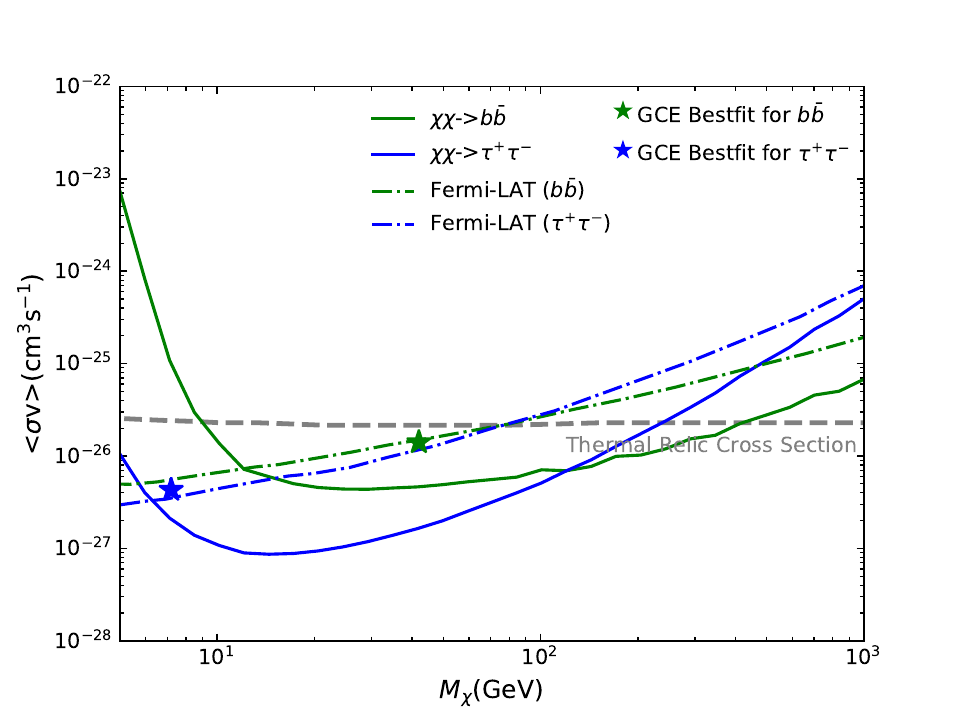}}
    \caption{{\it Left}--The 95\% C.L. upper limits on $\langle\sigma v\rangle$ as a function of $M_{\chi}$ for the annihilation channel $\mu^{+}\mu^{-}$. The solid black line shows the result with $D_{0}=10^{27}\,{\rm cm}^{2}\,{\rm s}^{-1}$, diffusion radius $r_{\rm h}=2\times r_{\rm max}$, and a uniform magnetic field $B_{0}=1\, \mu {\rm G}$. The dots line shows the median expected sensitivity while the green and yellow bands represent the $68\%$ and $95\%$ quantiles, derived from mock data. {\it Right}--The 95\% C.L. upper limits on $\langle\sigma v\rangle$ as a function of $M_{\chi}$ for the annihilation channels $\tau^{+}\tau^{-}$ (solid-blue) and $b\bar b$ (solid-green). In {\it both} panels, the stars represent the best-fitting value for GCE in the corresponding final states~\cite{DiMauro:2021qcf}. The dot-dash lines show the comparison with Fermi-LAT $\gamma$-ray search~\cite{Fermi-LAT:2015att}, and the dashed grey lines are the thermal relic cross-section value~\cite{Steigman:2012nb}.}
    \label{fig:mu}
\end{figure*}

\subsection{Results and discussions}

We now search for the DM contribution in the radio data and constrain the property of DM, by fitting 
\begin{equation}
    \chi^{2}(M_{\chi},\langle \sigma v \rangle)=\sum^{N}_{i}\frac{(S_{{\rm data},i}-S_{{\rm model},i})^{2}}{\sigma^{2}_{i}},
    \label{equ:chi2}
\end{equation}
where $S_{{\rm model},i}$ is the difference between the predicted flux density at the ON and OFF positions for a given $i$-th frequency, $S_{{\rm model},i}=S_{\nu_i}^{\rm ON}-S_{\nu_i}^{\rm OFF}$, $S_{{\rm data},i}$ is the observed flux density, $\sigma_{i}$ is standard deviations in each frequency bin obtained from the FAST data, and $N$ is the number of frequency bin left in Fig.~\ref{fig:Density_flux} . For $\mu^{+}\mu^{-}$, $\tau^{+}\tau^{-}$ and $b \bar b$ channels, we find no significant excess in the radio data, especially for DM mass favored by GCE as shown in Tab.~\ref{GCE best-fitting values}. Then we scan the DM mass for different annihilation channels to set upper limits at 95\% confidence level for $\langle \sigma v \rangle$ by requiring $\Delta \chi^{2}(\langle \sigma v \rangle)=\chi^{2}(\langle \sigma v \rangle)-\chi^{2}_{\rm min}=2.71$, where $\chi_{\rm min}^2$ is evaluated at the value of the cross-section minimizing the Eq.~(\ref{equ:chi2}) with specific DM mass.

For the $\mu^+\mu^-$ channel discussed in Ref.~\cite{DiMauro:2021qcf}, it could fit the GCE and survive from various constraints, derived from the combined dwarf spheroidal galaxies analysis and
the AMS-02 $\bar{p}$ and e$^+$ data analysis. Since this channel would generate a number of charged particles, our FAST observation of Coma Berenices could give very stringent constraints, as shown in the left panel of Fig.~\ref{fig:mu}. To assess the limits, we also generate 1000 sets of mock data with no DM contribution assumption for frequency bins in Fig.~\ref{fig:Density_flux} with the expected $\sigma_{{\rm T},i}$,  which is derived by Eq.~(\ref{equ:rms}) with the integration time the same as time bins after data cleaning, to get $\sigma_i$ for real data in each frequency. We then could get the expected limit bands by repeating the same analysis on these mock data\footnote{ We have also verified our analysis pipeline by successfully recovering an injected mock DM signal.}. Our limits derived from real data fall within one standard deviation
of the median expectation value, as shown in the left panel of Fig.~\ref{fig:mu}, suggesting our limits reasonable and consistent with no DM detection. Compared with constraints from the $\gamma$-ray observation of dSphs \cite{Fermi-LAT:2015att}, the constraints from FAST observation are strong enough to test the parameter space favored by the GCE \cite{DiMauro:2021qcf}, using our setup for the synchrotron emission.

In addition, we also show our results for $b \bar b$  and  $\tau^{+}\tau^{-}$ final states in the right panel of Fig.~\ref{fig:mu}. Similar to constraints derived from the  $\gamma$-ray observation of dSphs \cite{Fermi-LAT:2015att}, these upperlimits from the FAST observation are also in tension with parameters favored by the GCE \cite{DiMauro:2021qcf}. And for the mass of DM from about tens of GeV to several hundred GeV, the constraints from the FAST could be comparable and even stronger. 

\begin{figure*}
    \centerline{
    \includegraphics[width=1.\columnwidth]{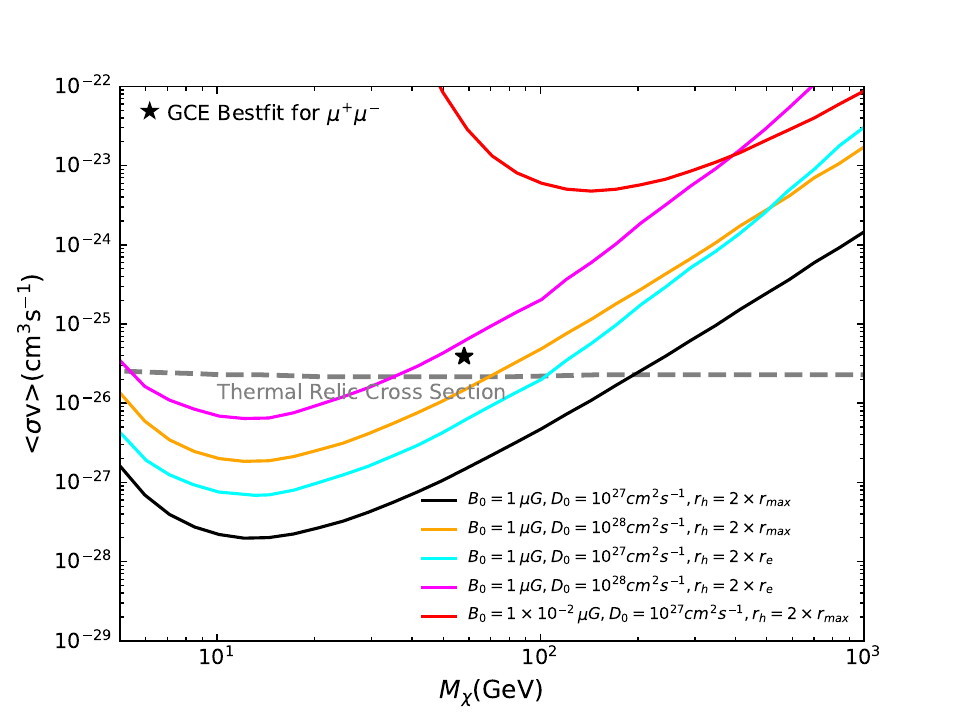}
    \includegraphics[width=1.\columnwidth]{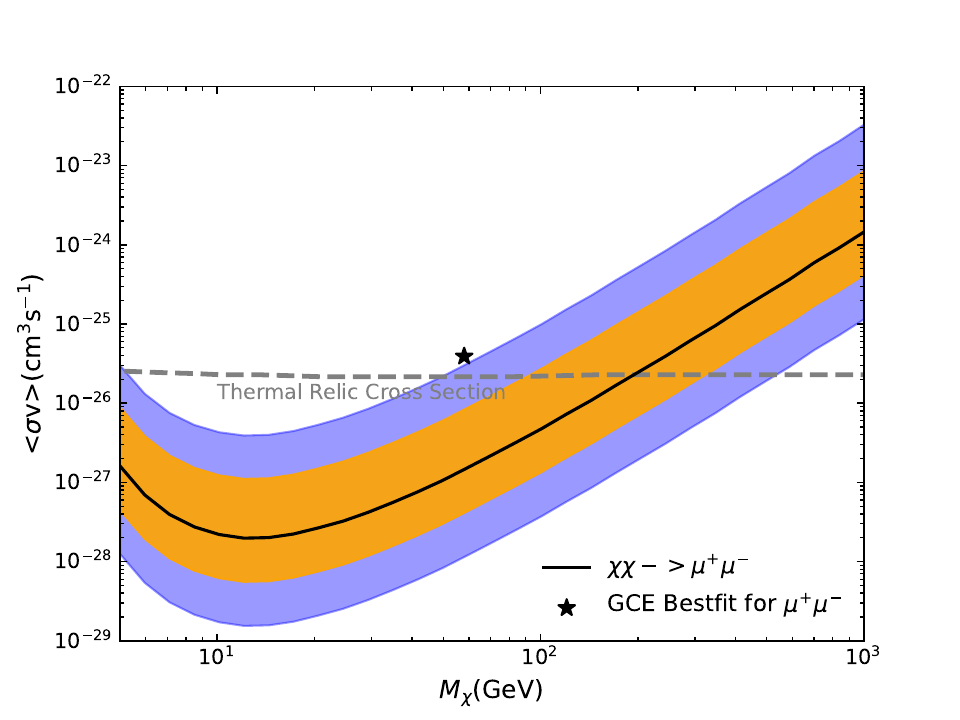}}
    \caption{{ The 95\% C.L. upper limits on $\langle\sigma v\rangle$ as a function of $M_{\chi}$ for the annihilation channel $\mu^{+}\mu^{-}$. {\it Left}--Lines with different colours represent the different values of the magnetic field $B_{0}$, diffusion constant $D_0$, and diffusion radius $r_{\rm h}$, as shown in the legend.}  { {\it Right}-- The solid black line represents the best-fitting halo parameters derived from the stellar kinematics~\cite{Bonnivard:2015xpq}. The orange and blue bands present the $68\%$ and $95\%$ confidence regions in the DM halo parameters space, which reflect the uncertainty in the DM halo parameters, using the method in Ref.~\cite{Natarajan:2013dsa}. In both panels, the black line is the benchmark case in this work, see Fig~\ref{fig:mu}.}}
    \label{fig:mu_discuss}
\end{figure*}

However, unlike the expected signal in $\gamma$-ray from prompt emission of DM annihilation, the expected radio emission associated DM annihilation depends on several astrophysical factors, which have no good measurements for dSphs unfortunately. After injection from DM annihilation, charged particles would diffuse in the dSphs. The diffusion coefficient, $D_0$, diffusion radius, $ r_{\rm h}$, and magnetic field, $B_{0}$, would shape the distribution of equilibrium electron density, and also affect the strength of the radio signal. The turbulent nature of the magnetic field leads to the diffusion process of charged particles.  For the MW,  the observation of the boron-to-carbon ratio could give an estimation for $D_0$, about $10^{28}\,{\rm cm}^{2}\,{\rm s}^{-1}$ \cite{Korsmeier:2016kha}. For dSphs, there is an estimation that  $D_0$ is possibly even as low as $10^{26}\,{\rm cm}^{2}\,{\rm s}^{-1}$ \cite{2018MNRAS.476.1756H}, but $D_0$ is usually used from $10^{26}\,{\rm cm}^{2}\,{\rm s}^{-1}$ to $10^{28}\,{\rm cm}^{2}\,{\rm s}^{-1}$ \cite{Colafrancesco:2006he,Spekkens:2013ik,Natarajan:2013dsa,Natarajan:2015hma, Colafrancesco:2014coa,Beck:2015rna,Regis:2017oet,Kar:2019hnj,Vollmann:2019boa, Bhattacharjee:2019jce, Bhattacharjee:2020phk, Kar:2020coz, Chen:2021rea, Regis:2021glv,Basu:2021zfg,Jeltema:2008ax,McDaniel:2017ppt}, considering its uncertainty. The diffusion radius, $r_{\rm h}$, determines the volume in which charged particles could be kept. In principle, the scale of $r_{\rm h}$ should depend on the distribution of the magnetic field, thus also depend on the distribution of luminous matter. Several times the size of the half-light radius, $r_{\rm e}$, and the distance of the outermost star of dShps, $r_{\max}$, are usually used to estimate $r_{\rm h}$ \cite{Colafrancesco:2006he,Vollmann:2019boa}, but  sometimes the size of DM halo is also used for the estimation \cite{Regis:2017oet}. The magnetic field, $B_{0}$, would affect the energy loss rate of charged particles and also the power of synchrotron emission. It was suggested that there is a magnetic field in the MW, which is not negligible even at the location of the dwarf galaxies in the neighborhood of the MW \cite{Regis:2014koa}. With the scaling relation and the distance, $d=44\,{\rm kpc}$, from the center of MW, Coma Berenices is bathed in a magnetic field of strength $B_{0}=1.14\,\mu{\rm G}$. On the other hand, previous star formation in dSphs may also leave relic magnetic fields in them, ranging from $0.4 \mu{\rm G}$ to several $\mu{\rm G}$~\cite{Regis:2014koa}. { A weak magnetic field in dSphs ($\sim 0.01\,\mathrm{\mu G}$) is also suggested, assuming the equipartition with the cosmic-ray density \cite{Regis:2014koa}.}  We set $D_0$, $r_{\rm h}$ and  $B_{0}$ to $10^{27}\,{\rm cm}^{2}\,{\rm s}^{-1}$, $476$ pc and 1 $\mu{\rm G}$ respectively, as our benchmark model. Increasing $D_0$ and decreasing $r_{\rm h}$ and $B_{0}$ would reduce the expected signal and weaken the constraints. As shown in  the left panel of Fig.~\ref{fig:mu_discuss}, the constraints for the $\mu^+\mu^-$ channel are still strong enough to test the DM scenario of the GCE, while we change $D_0$ to $10^{28}\,{\rm cm}^{2}\,{\rm s}^{-1}$, and switch  $r_{\rm h}=$ 148 pc to be twice the $r_{\rm e}$. { But varying these two factors simultaneously or decreasing $B_{0}$ to $0.01\,\mathrm{\mu G}$} would make constraints too weak to exclude the best-fitting point for the GCE.


{ Apart from these astrophysical factors, the DM distribution in the dSph can also be crucial for strengthening the expected DM annihilation signal. An astrophysical factor, commonly known as the $J$-factor and defined as $J=\int \rho_{\chi}^{2}{\rm d}l\,{\rm d}\Omega$, is widely used to quantify the strength of the annihilation signal. However, the expected synchrotron signal, depending also on the diffusion and energy loss, would not be linearly proportional to the $J$-factor, as in gamma-ray searches for prompt emissions from DM annihilation. Thus, we do not investigate the uncertainty of the $J$-factor, but instead investigating the uncertainty of DM halo parameters inferred from measurements of stellar kinematic~\cite{Bonnivard:2015xpq}, while testing the effect of the uncertainty of DM distribution on the derived constraints (see also, e.g. Ref.~\cite{Natarajan:2013dsa}). We determine the $68\%$ and $95\%$ confidence regions in the $\rho_{0}$-$r_{\rm s}$-$\alpha$ space and derive constraints by calculating synchrotron emissions with these DM halo parameters. In the right panel of Fig~\ref{fig:mu_discuss},  the solid black line is plotted for the best-fit halo parameters derived from stellar kinematics and the shaded regions show the uncertainty in the halo parameters. This indicates that the constraints on the DM can vary by approximately an order of magnitude, due to the uncertainty of DM halo parameters.}
{\footnotetext[2]{ https://github.com/alex-mcdaniel/RX-DMFIT}}

\begin{figure}
    \centering
    \includegraphics[width=1.\columnwidth]{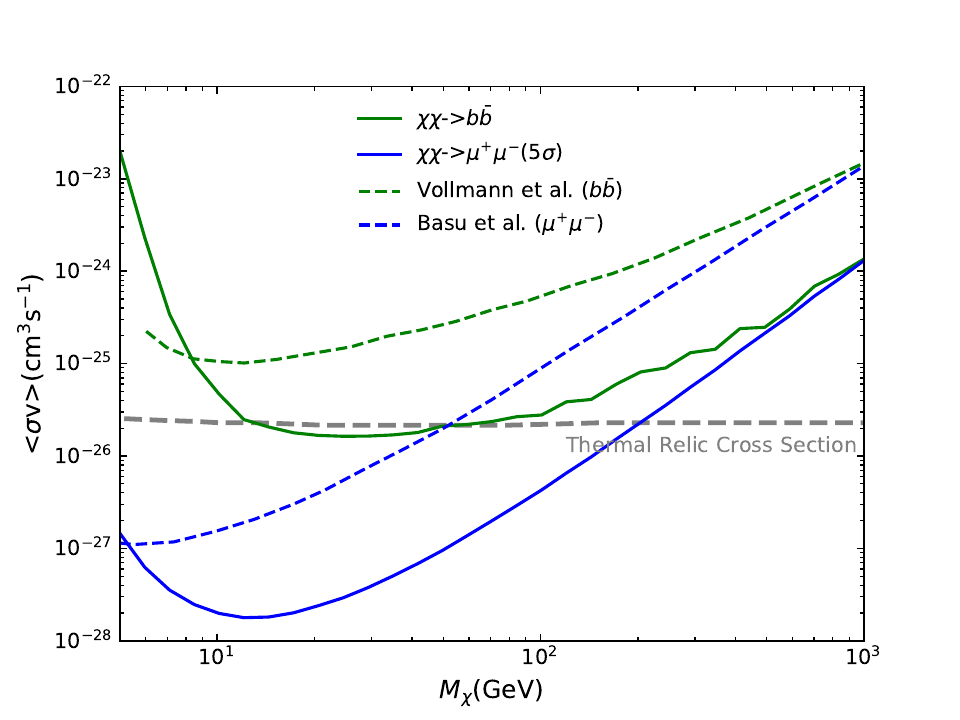}
    \caption{Constraints on $\langle \sigma v \rangle$ as a function of $M_{\chi}$ and comparison with other radio observations. The solid blue line shows the $5\sigma$ limit on pure $\mu^{+}\mu^{-}$ final state with diffusion constant $D_0=3\times10^{26}\,{\rm cm}^{2}\,{\rm s}^{-1}$, and about 11 times of Coma Berenices' core radius for diffusion. The green solid line shows the 95\% C.L. upper limit for $b\bar b$ channel with diffusion radius $r_{\rm h}=2\times r_{\rm e}$. The blue dashed line shows the comparison with radio observations from stacking analysis of 23 dSphs (TGSS) ~\cite{Basu:2021zfg}. The dashed green line shows the comparison with radio observations from Canes Venatici I (LOFAR) ~\cite{Vollmann:2019boa}. The thermal relic cross-section value is marked by the grey dashed lines~\cite{Steigman:2012nb}.}
    \label{fig:rsrhos_and_comparisons}
\end{figure}

To compare with other constraints derived from radio observation reasonably, we also set $r_{\rm h}=2r_{\rm e}$ as in Ref.~\cite{Vollmann:2019boa} to get the 95\% C.L. upper limits for the $b\bar{b}$ channel,  and set $D_0=3\times10^{26}\,{\rm cm}^{2}\,{\rm s}^{-1}$ as well as about 11 times of Coma Berenices' core radius for diffusion to get 5 $\sigma$ limits for the $\mu^+\mu^-$ channel as in Ref.~\cite{Basu:2021zfg}. {  
As shown in Fig.~\ref{fig:rsrhos_and_comparisons}}, the FAST observation could significantly improve constraints, with roughly an order of magnitude improvement compared with previous work, in a large range of mass, benefiting from the large size of the telescope.

\section{Conclusions}\label{conclusion}

The GCE, detected  by the Fermi-LAT, is an attractive but tentative signal from DM annihilation. After various sophisticated investigations, it is still difficult to discriminate the DM scenario from astrophysical scenarios solely by $\gamma$-ray observations. Thus, we use FAST observations to search for associated radio signals. We take the Coma Berenices dwarf galaxy as our target, and undertake two hours of ON/OFF observation.     
We find no significant continuum radio emission, which could be attributed to DM annihilation, from the location of Coma Berenices. Very stringent constraints are set for $\mu^+\mu^-$, $b\bar{b}$ and $\tau^+\tau^-$ final states. The FAST observation could improve constraints, compared with previous work using $\gamma$-ray and radio observations, in a large range of mass. The DM scenario for the GCE is in tension with the radio observation for reasonable choices of astrophysical factors. { However, considering the large uncertainty on astrophysical parameters, including the magnetic field, the diffusion coefficient, the diffusion radius, and the DM halo parameters, constraints derived from our observation could be weakened. Future observations to narrow the uncertainties of the astrophysical parameters can be helpful to  pin down the limits set by radio observation of dSphs,  such as an estimation of the line-of-sight integral of the magnetic fields in dSphs. This can be done by measuring the Faraday rotation of the polarization angle of the polarized emission from background galaxies along the line-of-sight, shown in Ref.~\cite{Gaensler:2005qj,Siffert:2010cc} for the LMC.} These results suggest that radio observation is a complementary way to search for DM, and further observations by FAST and other radio telescopes may test the DM scenario of the GCE reliably.

{ For the analysis in this work, we use the publicly available code {\tt RX-DMFIT}\footnotemark[2]. FAST dataset used in this work can be requested by contacting the corresponding author.} 

\begin{acknowledgments}
This work made use of the data from FAST (Five-hundred-meter Aperture Spherical radio Telescope).  FAST is a Chinese national mega-science facility, operated by the National Astronomical Observatories, Chinese Academy of Sciences.
YZM is supported by the National Research Foundation of South Africa under Grant No. 120385 and No. 120378, and National Natural Science Foundation of China with Project No. 12047503. X.H. is supported by the CAS Project 
for Young Scientists in Basic Research (No. YSBR-061), the Chinese Academy of Sciences, and the Program for Innovative Talents and Entrepreneur in Jiangsu.
YC acknowledges the support from the NSFC under grant No. 12050410259, and Center for Astronomical Mega-Science, Chinese Academy of Sciences, for the FAST distinguished young researcher fellowship (19-FAST-02), and MOST for the grant no. QNJ2021061003L. This work was part of the research programme ``New Insights into Astrophysics and Cosmology with Theoretical Models confronting Observational Data'' of the National Institute for Theoretical and Computational Sciences of South Africa. We thank Bo Zhang from the National Astronomical Observatories of the Chinese Academy of Sciences for her valuable comments and discussions.
\end{acknowledgments}


\end{document}